# Teaching Requirements Engineering for AI: A Goal-Oriented Approach in Software Engineering Courses

Beatriz Braga Batista
Federal University of Amazonas
Manaus, AM, Brazil
beatriz.souza@icomp.ufam.edu.br

Márcia Sampaio Lima
Amazonas State University
Manaus, AM, Brazil
msllima@uea.edu.br

Tayana Uchoa Conte
Federal University of Amazonas
Manaus, AM, Brazil
tayana@icomp.ufam.edu.br

## ABSTRACT

**Context**: Requirements Engineering for AI-based systems (RE4AI) presents unique challenges due to the inherent volatility and complexity of AI technologies, necessitating the development of specialized methodologies. It is crucial to prepare upcoming software engineers with the abilities to specify high-quality requirements for AI-based systems. **Goal**: This research aims to evaluate the effectiveness and applicability of Goal-Oriented Requirements Engineering (GORE), specifically the KAOS method, in facilitating requirements elicitation for AI-based systems within an educational context. **Method**: We conducted an empirical study in an introductory software engineering class, combining presentations, practical exercises, and a survey to assess students' experience using GORE. **Results**: The analysis revealed that GORE is particularly effective in capturing high-level requirements, such as user expectations and system necessity. However, it is less effective for detailed planning, such as ensuring privacy and handling errors. The majority of students were able to apply the KAOS methodology correctly or with minor inadequacies, indicating its usability and effectiveness in educational settings. Students identified several benefits of GORE, including its goal-oriented nature and structured approach, which facilitated the management of complex requirements. However, challenges such as determining goal refinement stopping criteria and managing diagram complexity were also noted. **Conclusion**: GORE shows significant potential for enhancing requirements elicitation in AI-based systems. While generally effective, the approach could benefit from additional support and resources to address identified challenges. These findings suggest that GORE can be a valuable tool in both educational and practical contexts, provided that enhancements are made to facilitate its application.

## 1 INTRODUCTION

As AI-based techniques become deeply embedded in software systems, revolutionizing performance and cost-efficiency, the traditional requirements engineering methods struggle to keep pace with the unique challenges posed by the evolving nature of AI development [2]. Requirements Engineering (RE) ensures that the developed system meets the stakeholders' needs and operates within the defined parameters [32]. However, the dynamic and complex nature of AI-based systems introduces significant challenges to the conventional RE methodologies [2].

AI systems add another layer of complexity due to their data-driven nature, continuous learning, and adaptability [22], which directly impact the quality of the resulting software. Poorly defined requirements for AI systems can result in unpredictable behaviors and hinder quality assurance efforts, making it difficult to align the final product with stakeholder expectations. Traditional systems operate with predefined rules and logic, whereas AI systems often involve machine learning models that evolve over time. This evolution can lead to unpredictable behaviors and outcomes, making it difficult to capture and specify requirements accurately [29]. Moreover, AI systems require consideration of non-functional requirements such as transparency, ethics, and fairness, which are not typically addressed in traditional RE [30]. Addressing these requirements thoroughly is crucial for maintaining software quality, as failure to do so can lead to ethical issues and degrade user trust in the system [23].

Given these complexities, it is imperative to equip future software engineers with the skills and methodologies necessary to effectively conduct RE for AI-based systems. Goal-Oriented Requirements Engineering (GORE) emerges as a promising approach to address these complexities [2]. GORE focuses on identifying and modeling the goals that a system needs to achieve, providing a structured way to capture both functional and non-functional requirements [27]. This approach helps in understanding the underlying motivations and broader objectives of the system, which is crucial for AI-based applications.

Research by Van Lamsweerde [27] has laid the groundwork for GORE by demonstrating its applicability in capturing high-level goals and refining them into detailed requirements. Further studies, have hinted the adaptation of GORE for AI systems, highlighting its potential to handle the specific challenges posed by AI [7].

Existing research has identified several benefits of GORE in AI-based systems, such as improved traceability, better management of conflicting requirements, and enhanced stakeholder communication [20]. Studies like those by Belani et al. [7] and Martínez-Fernández et al. [22] have contributed to the understanding of how GORE can be tailored to AI's needs, emphasizing the need for integrating human-centered guidelines and addressing ethical considerations. However, there is a gap in the literature regarding the practical implementation of GORE in educational settings and its effectiveness in teaching future software engineers how to elicit RE for AI-based systems.

To address these gaps, this research aims to explore two research questions: (RQ1) "How effective is GORE in facilitating the elicitation of requirements for AI-based systems?" and (RQ2) "What are the perceived benefits and challenges of using GORE in the context of AI-based systems according to undergraduate students?"

These questions guided an evaluation of GORE's applicability and effectiveness in an educational setting, providing insights into its practical implementation and showcasing the results of using GORE for AI system requirements elicitation.

The methodology involved an empirical study conducted with 34 undergraduate software engineering students. The study consisted of structured presentations, practical exercises, and surveys to assess students' experiences and perceptions. Students' performance in applying the KAOS approach was evaluated, and their elicited requirements were analyzed against key metrics of Human-Centered AI (HCAI). The results indicated a generally positive reception of the KAOS method, with 88% of students demonstrating proficiency. Survey responses highlighted the ease of learning GORE, with most students finding it moderately easy or easier to learn. The analysis revealed that GORE is effective in capturing high-level requirements but less so in detailed analysis and planning.

This empirical study demonstrates that GORE is a promising approach for teaching requirements engineering for AI-based systems. While the approach is generally well-received and effective in helping students capture fundamental requirements, there are areas for improvement that could enhance its ease of learning and comprehensiveness in an educational setting. The insights gained from this study provide a foundation for further research and development in the field of requirements engineering education, aiming to create more robust and adaptable methodologies that can support the evolving needs of AI-based system development and better prepare future software engineers.

This paper is structured as follows. Section 2 reviews the background information and existing literature on requirements engineering for AI-based systems and Goal-Oriented Requirements Engineering. Section 3 describes the methodology of the study, including the materials and procedures used. Section 4 presents the results, focusing on the effectiveness of GORE and the perceived benefits and challenges as identified by undergraduate students. Section 5 discusses the implications of the findings, highlighting the practical applications and challenges. Section 6 provides the conclusion, summarizing the key findings and their significance. Section 7 details data availability including a link to the artifacts of the research.

## 2 BACKGROUND AND RELATED WORK

This section delves into the current state of RE4AI, outlining the challenges and complexities these systems introduce compared to traditional software development. It further explores GORE as a potential solution to these challenges. By reviewing existing literature and identifying gaps, we underscore the necessity for further research, particularly in the context of eliciting RE4AI.

### 2.1 Requirements Engineering for AI-based systems

Requirements Engineering for AI-based (RE4AI) systems presents unique challenges and complexities that distinguish it from traditional software development [30]. The inherent volatility and complexity of RE are well-documented [14], with the field characterized by the involvement of interdisciplinary stakeholders and significant uncertainty[31]. As AI-based systems become more prevalent, the demand for high-quality software applications has increased. This places a critical emphasis on RE, which plays a vital role in addressing software quality characteristics [30].

Engineering software systems that incorporate AI components introduce new processes such as data management, model training, and design [8, 19]. In software engineering (SE), the AI code itself is relatively small compared to the overall process of building systems with AI components [24]. Building AI-based complex systems goes beyond merely using specific AI algorithms. The development itself becomes more complex as the required data and implemented algorithms become increasingly interdependent [7].

From the literature, it is evident that existing tools for traditional SE practices, particularly RE, cannot be appropriately utilized when building AI systems [2, 6, 9, 24, 30]. This misalignment necessitates the development of new RE techniques and tools tailored to the needs of AI systems.

Studies like Macedo et al. [13] highlight how important Software Requirements courses are in training future IT professionals. With the advent of AI and Machine Learning, new paradigms are introduced in the process of requirements elicitation, analysis, and documentation. The teaching of software requirements must adapt to the integration of these emerging technologies, ensuring that students are well-prepared to handle the complexities and dynamic nature of AI systems. As these technologies become more prevalent, there is a need for educational methodologies to evolve, incorporating advanced tools and techniques that reflect current industry standards and practices. This adaptation is essential for producing professionals capable of addressing the unique challenges presented by AI-based systems and for maintaining the quality and reliability of software applications.

Several challenges have been identified in the literature concerning RE4AI:

- Vague or High-Level Requirements: AI requirements are often vague or high-level. Given the black-box nature of most AI models, requirements engineers find it difficult to specify precise requirements for such systems, often resulting in requirements that are too high-level or vague [21, 22].
- Non-Functional Requirements (NFRs): Specifying and understanding NFRs in AI systems is particularly challenging. Certain NFR categories, such as fairness and transparency, hold more importance in AI systems compared to others, such as modularity in traditional systems [2].
- **Customer Expectations:** Handling customer expectations is another significant challenge. Organizations often do not realize that AI models are probabilistic and must learn patterns from messy data. This leads to difficulties in managing customer expectations regarding the limitations and performance of AI systems [11, 15, 16, 18].
- **Data Requirements:** Issues with data requirements, such as lack of structure, availability, or quality, pose significant challenges. Data is the cornerstone of AI systems, and poor data quality can severely impact the performance and reliability of the system [5, 9, 25].
- **Human-Centered Requirements:** AI systems often interact closely with human users, making it essential to incorporate user needs and feedback mechanisms into the RE process. Ensuring transparency and explainability helps build user trust and facilitates better interactions with the system [1].

A study [3] on human-centered AI systems identified six key areas of focus: user needs, model needs, data needs, feedback and user control, explainability and trust, and errors and failure management. These areas highlight the multifaceted nature of RE4AI, where technical performance must be balanced with human-centric considerations:

- **User Needs**: This area focuses on understanding and addressing the specific needs of users who interact with AI systems. It involves gathering user requirements, ensuring usability, and designing systems that meet user expectations and provide a positive user experience.
- **Model Needs**: This involves ensuring that the AI models used in systems are reliable, accurate, and appropriate for the intended tasks. It includes considerations for model selection, training, validation, and optimization to achieve high performance and robustness.
- **Data Needs**: The quality and availability of data are critical for training effective AI models. This area addresses the identification, collection, preprocessing, and management of data. It ensures that the data used is relevant, diverse, and free from biases to produce reliable AI outputs.
- **Feedback and User Control**: This area focuses on incorporating mechanisms for user feedback and control over the AI system. It includes designing interfaces and interactions that allow users to provide input, adjust settings, and influence the system's behavior, thereby increasing transparency and user trust.
- **Explainability and Trust**: Ensuring that AI systems are explainable and transparent is crucial for building user trust. This area involves developing methods and tools that help users understand how AI decisions are made, the rationale behind them, and how they align with human values and ethical standards.
- **Errors and Failure**: This area addresses the identification and management of potential errors and failures in AI systems. It involves designing systems that can detect, explain, and recover from errors, and ensuring that users are aware of the limitations and potential risks associated with AI systems.

There's a need for more empirical studies that have specifically tested the efficacy of methods for RE4AI, especially in the context of education. One promising approach is GORE. Studies suggest that GORE could be effectively adapted for AI systems like [1, 7, 17], but further analysis is needed to fully understand its applicability and to develop frameworks that can address the specific needs of RE4AI.

This motivated our research to analyze the efficacy, advantages, and disadvantages of using GORE for teaching requirements elicitation in AI systems. By exploring this approach, we aim to contribute to the development of robust RE methodologies that can support the dynamic and data-driven nature of AI systems. Our goal is to enhance the educational experience of future software engineers, preparing them to handle the complexities of AI systems.

### 2.2 Goal-Oriented Requirements Engineering

Developing AI components primarily involves applying techniques to achieve specific objectives. Despite this, there is a notable lack of validated techniques for addressing critical aspects of requirements engineering (RE) for AI systems [30]. Goals have long been a central concept in AI development and have proven effective in various applications [10].

Recognizing the need for a robust framework to tackle the unique challenges in RE for AI, researchers have identified the area of Goal-Oriented Requirements Engineering (GORE) as a promising starting point [17]. GORE has been adapted and applied to numerous subtopics within RE and beyond, including agent orientation, aspect orientation, business intelligence, model-driven development, and security [7].

GORE is an approach that focuses on capturing, modeling, and analyzing the goals that a system must achieve. It emphasizes the importance of understanding the underlying objectives of a system and refining these high-level goals into detailed requirements. The principles of GORE include identifying stakeholders' goals, decomposing these goals into sub-goals, and ensuring that all system requirements align with achieving these goals [27].

In a systematic mapping study conducted by Ahmad et al. [2], GORE emerged as the third most popular modeling notation and language used in RE studies. The study highlights the adaptability and broad application of GORE in addressing various RE challenges.

GORE offers several advantages, particularly in supporting non-functional requirements (NFRs) and business rules. It provides better mechanisms for modeling requirements at lower levels of abstraction compared to Unified Modeling Language (UML) [1]. According to Silva et al. [26], GORE can present requirements or concepts with fewer structural diagrams than UML, making it more concise and focused. However, GORE is generally more challenging to learn and is predominantly used by requirements engineers and software engineers. UML remains more widely known and utilized, especially among non-software engineers, due to its broader accessibility and familiarity.

One of the prominent methodologies within GORE is KAOS (Knowledge Acquisition in autOmated Specification [12] or Keep All Objects Satisfied [28]). KAOS provides a comprehensive approach to capturing both functional and non-functional requirements through goal models. Some advantages of using KAOS include its ability to systematically derive requirements from high-level goals, handle conflicting goals through obstacle analysis, and ensure traceability from goals to implementation [20].

Despite its potential, no recent studies have specifically addressed the application of GORE to the particular challenges of RE in AI-based systems, nor have there been empirical evaluations in an educational context to test its efficacy for teaching. This gap in the literature underscores the need for further research to validate and refine GORE approaches in the context of AI development. By addressing this gap, our research aims to provide evidence-based insights into the effectiveness of GORE in teaching RE4AI systems, which would inform the development of more robust and practical RE methodologies for AI. This will help future software engineers to be well-prepared to address these challenges.

## 3 METHODOLOGY

We conducted an empirical study in a software engineering class to evaluate the effectiveness of GORE and its application using the KAOS approach for AI-based systems. The study followed Wohlin's guidelines for empirical experimentation in software engineering [33]. In the Experiment Setup subsection, we will detail the steps taken to implement the study. The Evaluation Criteria subsection will describe how the evaluation was conducted, including the criteria used to assess the elicited requirements, the participants' application of the KAOS approach and their answers to the survey.

### 3.1 Study Setup

In this subsection, we describe the steps followed in our study to evaluate the teaching of GORE for AI-based systems. The procedure involves three main parts over the course of three days: Lecture, Training Session, and Data Collection.

*3.1.1 Lecture.* This initial session aimed to familiarize participants with the theoretical foundations necessary for the subsequent practical exercises. On the first day, the class received a lecture covering the principles of GORE, the KAOS approach and the six areas of HCAI [3], integrating it to the methodology aiming to provide a more comprehensive understanding of how it can address the unique challenges posed by AI systems. The participants also engaged in exercises designed to familiarize them with the concepts of HCAI and GORE. The material used will be available in Section 7. Following the presentation, the materials were made available to the participants for review and study at any time, ensuring they had continuous access to the information.

*3.1.2 Training Session.* On another day, we reviewed the GORE principles along with the HCAI areas and conducted a training session where participants practiced collectively eliciting the GORE and KAOS for an AI-based system. The objective was to provide participants with hands-on experience in applying the GORE concepts through the KAOS approach, reinforcing their understanding and preparing them for the practical exercise.

*3.1.3 Data Collection.* Next, the practical part of the study was conducted. On a subsequent class, the intentions of the experiment were explained in detail. Students were informed about the voluntary nature of their participation, the ability to withdraw without any consequences, and the confidentiality of their responses. After obtaining consent from those who agreed to participate, a simulated task was introduced. Students were provided with a paragraph describing an AI-based system along with a cheat sheet with KAOS elements and were asked to elicit the requirements for this system within around 100 minutes. The paragraph and cheat sheet used are also available in Section 7.

The requirements elicited by the participants during the task were collected and anonymized for further evaluation. To gather additional insights, we administered a survey following the elicitation task, with participants given 36 hours to respond. The survey collected demographic information and reflections on their experience using GORE for AI-based systems. Participation in the survey was voluntary, and only data from students who consented to the experiment were used for analysis. All collected data was anonymized and can be accessed in Section 7.

### 3.2 Evaluation Criteria

*3.2.1 Model Analysis.* To evaluate the elicited requirements, we proposed a checklist (Table 1) developed based on the six areas of HCAI and their most relevant aspects [3]. The checklist was reviewed by specialists through two rounds of discussion to ensure its accuracy and comprehensiveness. This checklist provided a systematic approach to evaluating the quality of the requirements elicited by the students.

The effectiveness was determined by the percentage of students who successfully addressed the criteria listed in the checklist. For each area, an "Overall" effectiveness score was calculated as the mean of the effectiveness percentages of the individual criteria within that area.

Additionally, the students' application of the KAOS approach was assessed and categorized on a scale from 1 to 5, each assessment was reviewed by a specialist:

(1) **No Application**: The student did not utilize the KAOS approach in their requirement elicitation task.
(2) **Incorrect Application**: The student confused goals with requirements, indicating a fundamental misunderstanding. Specifically, they treated requirements as non-leaf elements rather than properly distinguishing them from higher-level goals.
(3) **Partial Understanding**: The student demonstrated a partial understanding of the KAOS approach but failed to appropriately apply multiple key elements or links. Their work reflected an initial attempt but lacked completeness or correctness.
(4) **Minor Inadequacies**: The student showed an adequate understanding of KAOS with minor inadequacies in the use of elements or links, indicating a generally good comprehension of the approach.
(5) **Correct Application**: The student accurately and effectively applied the KAOS approach. This included correctly distinguishing between goals and requirements, appropriately using elements and links, and providing a comprehensive and accurate set of elicited requirements.

Each student's performance was systematically evaluated based on these criteria to provide a quantitative measure of their proficiency in applying KAOS. This evaluation aimed to assess the overall ease of use in facilitating requirements elicitation for AI-based systems. Only those with minor inadequacies and correct applications according to the Likert Scale above had their requirements analyzed.

After conducting the initial assessment, the results were reviewed and discussed with a requirements engineering specialist during an additional round of review. This collaborative review process enhanced the thoroughness and reliability of the evaluation.

*3.2.2 Personal Opinion Survey.* The survey collected both quantitative and qualitative data on various aspects of the approach, including ease of learning, perceived usefulness, and specific challenges faced.

The quantitative data from the survey were analyzed using descriptive statistics. The distribution of responses to Likert scale questions was examined to identify overall trends and any significant variations in perceptions. Qualitative data from open-ended questions were grouped into key concepts to identify common patterns and insights.

Data collected from the requirement elicitation task and the survey responses were analyzed to determine the effectiveness of GORE in facilitating the elicitation of requirements for AI-based systems and to identify the perceived benefits and limitations of using GORE, as reported by undergraduate students. The checklist results provided quantitative measures of the quality and comprehensiveness of the elicited requirements, while the survey responses offered qualitative insights into the students' experiences and perceptions.

## 4 RESULTS

To evaluate the applicability and effectiveness of the GORE approach in teaching the elicitation of requirements for AI-based systems, we conducted a study involving undergraduate students. Out of the 34 participants, 30 completed the survey, resulting in an 88.2% response rate.

In subsection 4.1, we present and analyze the background of the participants in terms of their prior exposure to Requirements Engineering and AI concepts. In subsection 4.2, we discuss the proficiency levels of participants in applying the KAOS approach, including their perceptions of the ease of learning the approach. In subsection 4.3, we evaluate the quality and comprehensiveness of the requirements elicited by the participants. Subsections 4.4 to 4.7 present insights from the survey responses on various aspects of using GORE for AI-based systems.

### 4.1 Participant characterization

Concerning participant characterization, Figure 1a shows that the majority (70%) of the participants had only been exposed to Requirements Engineering through the coursework in the software engineering class. While Figure 1b shows that most participants (56,7%) had studied AI concepts as part of their curriculum, they lacked practical experience in applying these concepts. Importantly, none of the participants had prior experience with GORE before this study.

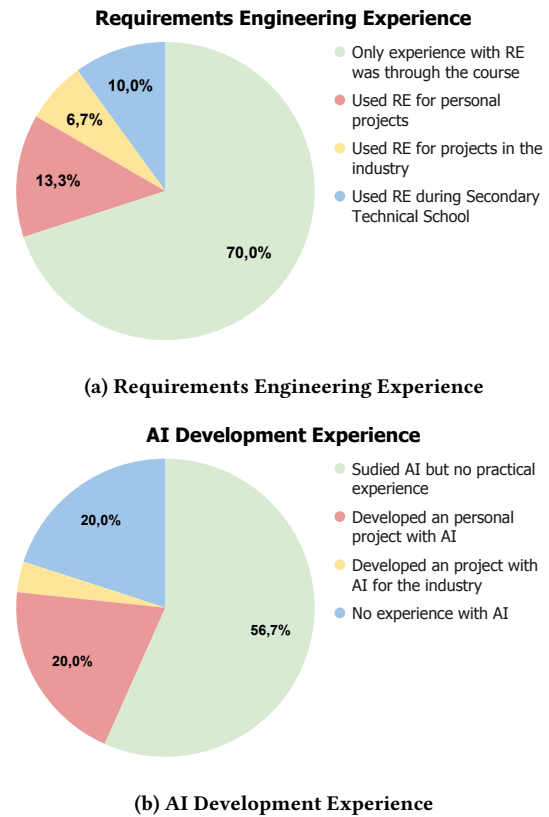

(a) Requirements Engineering Experience

(b) AI Development Experience

Figure 1: Participants' characterization

### 4.2 Ease of Learning

The evaluation of the exercises revealed varying degrees of proficiency among the participants in applying the KAOS approach which can be seen in Figure 2.

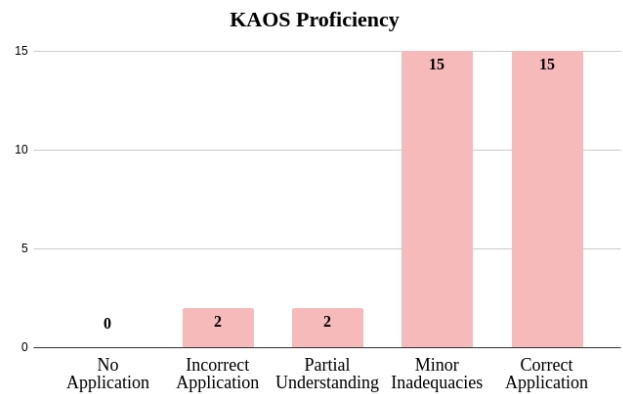

Figure 2: Participants' KAOS proficiency

According to the Likert scale established in the methodology, where 5 indicates "Correct Application" and 4 indicates "Minor Inadequacies," a total of 30 students (88%) either applied KAOS correctly or with minor issues. This percentage indicates that the majority of participants were able to grasp and apply the approach effectively after being introduced to it through the presentation and exercises. Only 4 participants (12%) had an incorrect application (Degree 2 on the scale) or partial understanding (Degree 3 on the scale).

Table 1: Checklist Used for Evaluation

| Area | Code | Comments |
| --- | --- | --- |
| User Needs | UN1 | Have the students clearly outlined what the user expects from the system? |
|  | UN2 | Have the students explained why the system is needed? |
|  | UN3 | Have the students identified all potential users of the system? |
| Model Needs | MN1 | Have the students decided between static vs. dynamic training? |
|  | MN2 | Have the students specified how the algorithm will be optimized? |
|  | MN3 | Have the students defined the tasks the model will be used for? |
| Data Needs | DN1 | Have the students identified the data sources and datasets that will be used? |
|  | DN2 | Have the students specified the features that will be used in the model? |
|  | DN3 | Have the students explained how the data will be labeled? |
| Feedback & Control | FC1 | Have the students ensured privacy measures in collecting feedback? |
|  | FC2 | Have the students planned how feedback will be collected, e.g., through surveys? |
|  | FC3 | Have the students specified when and how users should take control of the system? |
| Explainability & Trust | ET1 | Have the students explained how the user's information will be used? |
|  | ET2 | Have the students explained the system functionalities? |
|  | ET3 | Have the students explained how the system's output will be presented to the user? |
| Errors & Failure | EF1 | Have the students specified potential sources of errors? |
|  | EF2 | Have the students detailed possible errors that might occur? |
|  | EF3 | Have the students planned how to handle feedback from rejected predictions? |

The survey results provide further insights into the participants' perceptions of the ease of learning GORE/KAOS. When asked to rate the difficulty of learning the approach on a scale from 1 to 5 (with 1 being extremely difficult and 5 being extremely easy), the responses were distributed as seen in Figure 3.

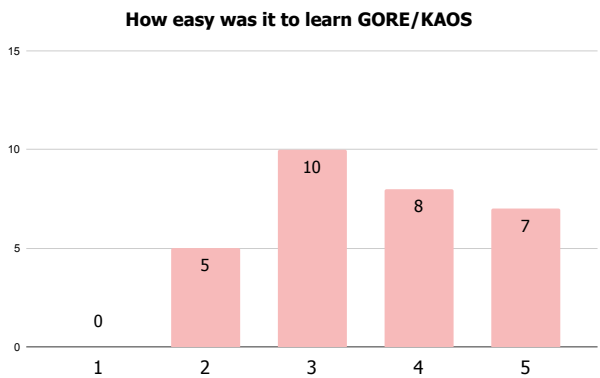

Figure 3: Participants' answers when asked how easy was it to learn GORE/KAOS

No participant rated the learning difficulty as 1, indicating that none of the participants found the approach to be extremely difficult to learn. The majority of the participants rated the ease of learning GORE/KAOS as either 3 or higher, with the most common rating being 3 (33.3%), followed by 4 (26.7%) and 5 (23.3%). This suggests that, overall, participants found GORE/KAOS to be moderately easy to learn, with a significant portion finding it relatively straightforward (ratings 4 and 5).

The survey responses reflect a generally positive perception of the ease of learning GORE/KAOS, with the majority (83%) rating it as moderately easy (3) or easier (4 and 5). The absence of ratings at 1 suggests that the initial instructional materials and exercises were effective in making the approach accessible to the participants.

### 4.3 Effectiveness

In evaluating the requirements elicited by the participants, the analysis excluded the 4 participants who demonstrated incorrect application or partial understanding of the KAOS approach, focusing only on the 30 participants who showed minor inadequacies or correct application. The results are summarized in Table 2.

The overall effectiveness of GORE in facilitating comprehensive requirements elicitation can be observed through the varied levels of completion across different categories.

- **User Needs**: The high percentage for UN1 and UN2 indicates that all participants managed to capture basic user expectations and the necessity of the system. However, the low percentage for UN3 highlights a gap in identifying all potential users, suggesting a need for more focus on comprehensive user analysis.
- **Model Needs**: The results indicate a strong performance in defining model needs, particularly in deciding training

Table 2: GORE Effectiveness

| Area | Code | Effectiveness | Overall |
|---|---|---|---|
| User Needs | UN1 | 100.00% | |
| | UN2 | 100.00% | 73,33% |
| | UN3 | 20.00% | |
| Model Needs | MN1 | 96.67% | |
| | MN2 | 70,00% | 83,33% |
| | MN3 | 83,33% | |
| Data Needs | DN1 | 93,33% | |
| | DN2 | 50,00% | 72,22% |
| | DN3 | 73,33% | |
| Feedback & Control | FC1 | 50,00% | |
| | FC2 | 96,67% | 60,00% |
| | FC3 | 33,33% | |
| Explicability & Trust | ET1 | 53,33% | |
| | ET2 | 43,33% | 43,33% |
| | ET3 | 33,33% | |
| Errors & Failure | EF1 | 53,33% | |
| | EF2 | 43,33% | 41,11% |
| | EF3 | 40,00% | |

methods and defining tasks. However, optimization methods were less frequently specified, indicating an area for improvement.
- **Data Needs**: Participants performed well in identifying data sources and labeling methods but were less thorough in specifying features, suggesting a partial understanding of detailed data requirements.
- **Explainability & Trust**: The high percentage for planning feedback collection indicates that participants understood the importance of gathering user input. However, ensuring privacy measures and defining user control mechanisms were less frequently addressed, indicating significant areas for improvement.
- **Explainability & Trust**: The relatively low percentages in this area highlight a clear need for better integration of explainability and trust in requirements elicitation, which are crucial for user acceptance of AI systems.
- **Errors & Failure**: The results show a substantial gap in addressing errors and failure, indicating that participants need more guidance on identifying and planning for potential issues in AI systems.

Overall, the evaluation of the elicited requirements shows that GORE is effective in helping participants capture fundamental user and model needs, as evidenced by percentages in areas such as defining user expectations (100%), system necessity (100%), and model training (96.67%) decisions. However, the approach appears less effective in areas requiring detailed analysis and planning, such as identifying all potential users (20%), specifying features (50%), ensuring privacy (50%), and handling errors (40%).

## 4.4 Participants' Perception on GORE's Adaptability for AI

The survey responses on the ease of defining requirements with GORE provide strong evidence of its perceived benefits. When asked "Using GORE makes it easier to define requirements for AI-based systems" on a scale from 1 (Strongly Disagree) to 5 (Strongly Agree), the responses can be seen in Figure 4.

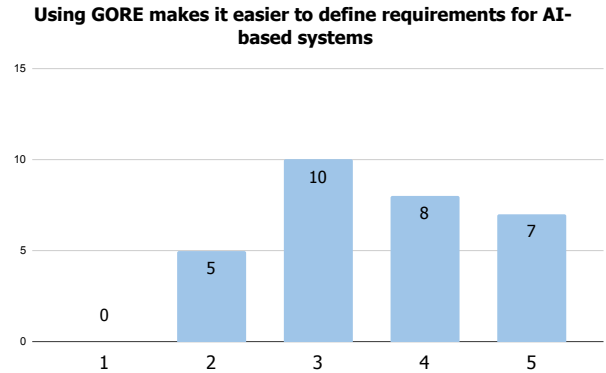

Figure 4: Participants' opinions when asked if they agree that using GORE makes it easier to define requirements for AI-based systems

The percentage of 50% being strong positive responses (ratings of 4 and 5) with highlight the effectiveness of GORE in making the requirements elicitation process easier for AI-based systems. The percentage of 17% negative feedback (rating of 2 and 1) indicates that the GORE approach does not pose significant barriers to application. This is encouraging, as it suggests that any GORE has the potential to support requirements for AI-based systems.

## 4.5 Perceived Benefits

To assess the participants' insights into the specific benefits perceived by the participants, we asked the following question:

- Which aspects of GORE did you think were especially useful for the requirements elicitation for AI-based systems?

These benefits are categorized into key themes including:

**Goal Orientation**: Half of the participants (50%) appreciated the goal-oriented nature of GORE, which helps in defining and decomposing objectives, leading to a greater focus on specific aspects of development. This approach enables a broader understanding and traceability of system requirements. For example, P1: *"Defining and decomposing objectives leads to greater focus on certain aspects of development that might otherwise go unnoticed, such as data diversification."*, P13: *"The possibility of unraveling the objectives before moving on to the requirements themselves."*

**Practical Implementation**: The practical aspects of implementing the KAOS approach were highlighted as a significant benefit. Participants found the approach straightforward and useful in practice. P27: *"Practicality in implementing the KAOS approach."*.

**Structured Approach**: Participants found the structured approach of GORE beneficial, particularly the use of grouped diagrams and the direct approach of goal visualization. This structured representation facilitated a clearer understanding of system functionalities and requirements. P21: *"The grouped diagrams were very useful for final understanding of the requirements."*, P24: *"The most direct approach, based on goals that are easy to visualize and the way of asking the reason for the system's functionalities."*, P26: *"The most direct approach, based on goals that are easy to visualize and the way of asking the reason for the system's functionalities."*.

**System vs. User Perspective**: GORE allows for a more comprehensive view of the system's internal workings, which is crucial for AI systems. This system-centric perspective is often missing in traditional methods that focus heavily on user interactions. P11: *"I believe it is possible to see more of the system's point of view. In other methods we see more of the user's point of view. In the case of AI, specifically, many things happen 'behind the scenes,' without user interaction or knowledge. That's why I think it's more difficult to elicit in the traditional approach for AI."*.

**Added Value Perception**: The focus on objectives helps users perceive the added value of the system, leading to the elicitation of more fundamental and important requirements. P7: *"The fact that it is based on the objective makes it easier for users to perceive the added value, which allows them to elicit more important and fundamental requirements."*.

**Not Specified Order**: Some participants found the lack of a specific order in GORE beneficial, as it allows for a more flexible listing of ideas. P14: *"Because it doesn't follow an order, I think it's interesting to just list the ideas."*.

**Subdividing the Problem**: The ability to subdivide problems into smaller, manageable parts was seen as a valuable feature of GORE, aiding in a clearer understanding and resolution of complex issues. For example, P15: *"The fact of breaking the goals into sub-goals to be able to better understand and solve smaller problems."*

Additionally, the other participants either reported no benefits or mentioned the six areas of HCAI as benefits.

### 4.6 Perceived Challenges

To assess the participants' perception of the challenges faced by participants while using the GORE and KAOS, we asked the following question.

- What aspects of GORE did you find especially problematic for requirements elicitation for AI-based systems?

These challenges are categorized into key themes including:

**How to Refine or Stop Refining Goals**: Participants expressed difficulties in determining how to effectively refine goals and when to stop the refinement process. This issue highlights a lack of clear stopping criteria for goal refinement, leading to confusion about how specific goals should be before transitioning them into requirements. P18: *"You start with very vague goals and there comes a time when you don't know how to create a requirement based on that."*, P25: *"To what extent should I unravel? The lack of a well-established stopping criterion."*

**Diagram Complexity**: Participants found the complexity of assembling and managing the GORE diagrams to be problematic. The diagrams often became large and cluttered, making it difficult to maintain clarity and organization. The visual representation can become overwhelming due to the amount of information. P5: *"Assemble the diagram"*, P9: *"The diagram may end up being very large, which pollutes the analysis a little.*, P14: *"The amount of information that is piled up"*

**Lack of Specific Order**: The absence of a specific order or sequence in specifying goals during the analysis phase was confusing for some participants. This lack of structured progression made it challenging to follow a logical flow in goal specification. P8: *"Not following an order during the analysis. It confused me in some aspects"*, P24: *"It does not have a specific order of specifying these goals."*

**Prior Knowledge Required**: Participants noted that successfully applying GORE often required significant prior knowledge and experience working on the system. This requirement made it difficult for those who lacked a deep understanding of the system being analyzed. P4: *"Analyze which requirements can be included in the objectives and the prior analysis of multiple requirements possibilities. It is different from conventional requirements development."*, P26: *"Dividing sub-problems itself depends largely on previous experience in simulating a possible process"*, P30: *"A lot of prior knowledge about the basis of the system is required. Depending on the level of knowledge, it may be more difficult to visualize the goals."*

**GORE Limitations**: Participants pointed out limitations inherent to GORE. Issues such as the inability to link a requirement or objective to multiple parent objectives and the confusion caused by only allowing leaf-goals to be requirements were highlighted as significant drawbacks. P1: *"[...] I found the inability to link a requirement or objective to more than one parent objective very frustrating. There are many situations where a requirement is necessary for multiple objectives, but I could not normally describe this in the diagram, [...]"*, P10: *"I think knowing how to combine the leaves in each part"*, P15: *"The fact of addressing the 6 areas separately is confusing when creating sub-goals as one ends up being linked to several others, which pollutes the visual representation of GORE."*, P23: *"Only the leaves are requirements"*

**Difficulty in Understanding GORE or KAOS**: Participants faced challenges in understanding some of the fundamental concepts and syntax of GORE or KAOS. Specific elements like arrows and their conditions were confusing, and overall comprehension of the domain and requirements posed difficulties for some participants. P6: *"Domain and requirements"*, P13: *"Syntax and organization"*, P20: *"I don't know why all arrows have to have a ball if there are arrows that are neither or or and, it's an arrow without a special condition but it has to have a ball, I don't understand."*, P27: *"Understanding some concepts"*. P21 also answered the question with misconceptions about the approach, evidencing once again the lack of understanding of the methodology.

Notably, participants P2, P3, and P13 mentioned challenges related to AI areas rather than GORE itself. Additionally, the other participants reported no complaints, indicating that a portion of the class did not experience major issues with GORE.

### 4.7 Participants' General Opinion on GORE for AI

The survey responses to the question "What is your general opinion on using GORE for requirements elicitation for AI-based systems projects?" provide a comprehensive overview of participants' perceptions and experiences with the GORE approach. These responses highlight various aspects, including its effectiveness, ease of learning, and potential areas for improvement.

- **Overall Positive Perception**: Participants expressed a positive overall opinion of GORE. These responses suggest that GORE is well-received and considered a valuable tool for requirements elicitation. P22: *"So far the best approach I've seen to elicit requirements for AI systems."*, P11: *"I thought it performed better than the traditional method."*
- **Effectiveness in Managing Complexity**: Participants highlighted GORE's usefulness in organizing and eliciting requirements for complex and unintuitive systems. P2: *"Useful for organizing and eliciting the requirements of a system that is still complex and unintuitive,"*, P18: *"It seems like a form of Brainstorming, maybe it helps when someone has no idea where to start. [...]"*
- **Dynamic Requirements Gathering**: GORE was praised for its effectiveness in dynamic requirements gathering. This highlights GORE's strength in adapting to evolving requirements, though it also points out the challenge of managing large quantities of requirements. P9: *"I found it quite effective for dynamic requirements gathering. However, it can make analysis difficult if requirements accumulate in large quantities."*
- **Visual and Organizational Clarity**: The visual and organizational aspects of GORE were appreciated. These responses highlight the value of GORE's structured and visual approach. P21: *"The grouped diagrams were very useful for final understanding of the requirements,"* and another mentioned, *"The issue of having a way to represent or and is also interesting, in addition to the representation of inheritance being very intuitive."*
- **Challenges and Areas for Improvement**: Despite the positive feedback, participants also identified several challenges. P13: *"I think it could be the ideal model for this use, but it would need some adaptations."*. P12: *"[...] In my opinion, there is room for adapting the approach to make learning and implementation a little easier."*
- **Educational Value**: Participants appreciated GORE's educational value in understanding AI system requirements. P10: *"I found it very interesting. I believe it is a technique that helps a lot in this specific aspect, since, in my experience, generally when we create an AI model these things are not thought of."*

## 5 DISCUSSION

The results of this empirical study provide valuable insights into the applicability and effectiveness of Goal-Oriented Requirements Engineering (GORE) for AI-based systems.

In Subsection 5.1, we answer our RQ1 analyzing the elicited requirements, revealing GORE's strengths in capturing fundamental user and model needs, while identifying areas needing additional support for detailed analysis and planning. In Subsection 5.2, we answer our RQ2 discussing the ease of learning, benefits, and challenges perceived by the participants, providing a comprehensive overview of their experiences and feedback. Subsection 5.3 explores the broader implications of our findings for educational settings in requirements engineering for AI-based systems.

### 5.1 How effective is GORE in facilitating the elicitation of requirements for AI-based systems?

The analysis of the elicited requirements revealed that GORE is particularly effective in capturing fundamental user and model needs. The completion rates (Table 2) in categories such as defining user expectations, system necessity, and model training decisions highlight GORE's strength in addressing high-level requirements. However, the lower percentages in areas like specifying features, ensuring privacy, and handling errors indicate that the approach is less effective in facilitating detailed analysis and planning.

These findings suggest that while GORE provides a robust framework for high-level requirements elicitation, additional emphasis and support are needed to enhance its effectiveness in more complex areas. This could involve integrating complementary techniques or tools to address detailed requirements more comprehensively.

### 5.2 What are the perceived benefits and challenges of using GORE in the context of AI-based systems according to undergraduate students?

*5.2.1 Ease of learning.* The results from both the exercise evaluations and the survey responses indicate a generally favorable view of GORE/KAOS among the participants. Despite the lack of prior experience with GORE, the participants were largely able to grasp and apply the approach within the context of the study. The fact that nearly half of the participants (44%) applied the approach correctly and another portion (44%) had only minor inadequacies suggests that GORE/KAOS can be effectively taught and learned within a relatively short period.

The positive survey responses further support this, showing that a majority (83%) of participants found the approach easy to learn. These findings highlight the potential of GORE/KAOS as a viable approach for requirements engineering in educational settings, particularly for AI-based systems. However, the presence of incorrect applications and partial understandings indicates that additional support and practice may be needed for a subset of participants to fully master the approach.

According to Ahmad et al. [2], GORE can be challenging to learn, especially for non-software engineers [26]. Despite these challenges, the participants in our study were largely able to grasp and apply the approach within the context of the study.

*5.2.2 Benefits.* Participants identified several benefits of GORE, including its goal-oriented nature, structured approach, and ability to manage complex requirements. The emphasis on defining and decomposing objectives was particularly appreciated, as it enhanced understanding and traceability of system requirements. This structured method was seen as helpful in organizing and visualizing requirements, which is crucial for AI-based systems with complex, interdependent components.

GORE provides better support for NFRs and business rules compared to traditional RE methods [26], enhances communication between developers and stakeholders, and offers traceability [27]. The study's findings confirm that GORE is well-suited for requirements elicitation in AI-based systems, particularly due to its structured, goal-oriented approach, which helps break down complex requirements into manageable components—crucial for AI systems. Participants' positive feedback on GORE's ability to handle dynamic and evolving requirements further highlights its suitability for AI-based projects.

*5.2.3* **Challenges**. Despite the overall positive perception, several challenges were identified. Participants found it difficult to determine how to refine goals and when to stop the refinement process, indicating a need for clearer stopping criteria. The complexity of GORE diagrams and the lack of a specific order in goal specification also posed difficulties, suggesting that additional tools or guidelines to manage diagram complexity and provide a more structured progression could be beneficial. Some problems like uncertainty about the correctness of the decomposition and difficulties with internal completeness check also appeared in [4].

The inherent limitations of GORE, such as the inability to link a requirement to multiple parent objectives, highlight areas where the approach could be improved. Addressing these challenges could involve developing more intuitive and flexible diagramming tools to continuously refine the approach based on user experiences.

## 5.3 Educational and Practical Implications

Despite challenges, the reception of GORE among participants highlights its potential as a valuable educational tool. In an academic context, GORE helps students manage the complexities of AI systems by effectively capturing user (73.33%), model (83.33%), and data needs (72.22%). Its structured, goal-oriented approach and visual clarity make it useful for teaching requirements engineering and system analysis.

Practically, the use of GORE can enhance the requirements elicitation process by providing a clear, systematic approach for defining and decomposing goals. This is particularly beneficial for AI-based systems, where traditional requirements engineering techniques may fall short. The ability of GORE to manage complexity and provide a comprehensive view of system requirements can lead to more robust and well-defined specifications, ultimately contributing to the development of higher-quality AI systems.

However, the study also indicates that additional support, such as more detailed guidelines for goal refinement, clearer stopping criteria, could further enhance the effectiveness and ease of learning of GORE. By addressing these areas for improvement, educators and practitioners can better leverage GORE's strengths and mitigate its challenges, leading to more successful outcomes in a learning environment application.

## 6 CONCLUSION

The empirical study demonstrates that GORE, particularly through the KAOS approach, is a promising method for teaching requirements engineering for AI-based systems. The approach is generally well-received and effective in helping students capture fundamental requirements. However, there are areas for improvement that could enhance its ease of learning and comprehensiveness in educational settings.

This study contributes to requirements engineering education by providing evidence that GORE can be effectively integrated into the curriculum for eliciting AI-based system requirements. Its structured, goal-oriented nature helps students manage the complexities of AI systems, improving understanding and traceability of requirements. These findings align with literature, such as [2], which highlights GORE's effectiveness in capturing both functional and non-functional requirements.

Despite these positive outcomes, the study identified some limitations. The sample size was small, with only 34 participants, which may limit the generalizability of the findings. Additionally, the study took place in a controlled educational environment, which may not fully represent the complexities of real-world AI projects. The learning curve associated with GORE, as noted in the literature [26], suggests that further support and practice are needed for students to master the approach.

Requirements Engineering for AI is crucial for software quality. Pargaonkar [23] notes that integrating requirements engineering with quality assurance ensures high-quality software. Clear requirements support efficient testing, reduce rework, and strengthen quality assurance. Likewise, robust quality assurance ensures software meets its requirements, reducing the gap between user expectations and the final product. Given AI systems' complexity, RE4AI addresses both functional and non-functional requirements [7], making it essential to teach RE4AI to prepare engineers for developing high-quality AI solutions.

Future research should address these limitations by conducting studies with larger, more diverse samples and exploring GORE's application in real-world AI projects. Investigating the long-term retention of GORE concepts and their practical use in professional settings would provide valuable insights into its effectiveness and adaptability.

The insights from this study provide a foundation for further research and development in requirements engineering education. By creating more robust, adaptable approaches, educators can better support the evolving needs of AI-based system development and better prepare future software engineers.

## 7 DATA AVAILABILITY

The data collected and material used in research will be available at: https://anonymous.4open.science/r/SBQS_2024-D389/.

To avoid leakage of sensitive data and ensure privacy, we choose to anonymize all personal information provided in this paper.

## ACKNOWLEDGMENT

We thank all the participants in the empirical study, USES Research Group members for their support and the Research and Development (R&D) project 001/2020, signed with Federal University of Amazonas and FAEPI, Brazil, which has funding from Samsung, using resources from the Informatics Law for the Western Amazon (Federal Law nº 8.387/1991), and its disclosure is in accordance with article 39 of Decree No. 10.521/2020. We would like to thank the financial support granted by CNPq 314797/2023-8.